\documentstyle[11pt]{article}

\begin{document}

\hoffset=-20mm
\voffset=-8pt

\vspace*{26mm}

{\Large\bf
\noindent
On the Astronomical Records and Babylonian\\[3mm]
Chronology}\\[6mm]
{\large V.G.Gurzadyan}\\[6mm]
{ICRA,  University  of  Rome ``La Sapienza", Italy  and   Yerevan  Physics
Institute, Armenia}\\[5mm]

(Published in {\it Akkadica},  v.119-120 (2000), pp. 175-184.)

\section{METHODOLOGY}

This paper, which covers some aspects of the astronomical background of the 
chronology proposed by Gasche et al. (1998), is offered here for two reasons.  
That brief volume was intended to present a new approach and was not regarded 
as a comprehensive treatment of chronological issues in the ancient Near East, and 
certainly not as the final word on the role astronomy can play in solving 
chronological problems.  It was to be hoped that polemical debate could be 
avoided and the ensuing critical discussion would be based on the data and 
methodology presented.  The drawbacks of the approach are now apparent, however.  
On the one hand, there appears to be a real need (e.g., Huber 1999/2000b) for some
 additional general explanation of the astronomical conclusions published in that 
volume.  On the other hand, it is increasingly evident that the work of colleagues 
(particularly Huber) who have employed astronomical methods requires a more 
direct kind of attention.  The little attention payed to Huber and the lack of 
astronomical detail in Gasche et al. (1998) can be attributed to the fact that the
 book's origins lay in the study of the archaeological material and the Middle 
Chronology, and thus neither astronomy nor the High Chronology were at the 
center of the project.  A discussion of the archaeology must be left to my colleagues, 
but papers by Huber (1999/2000a, 1999/2000b, and Huber, this volume) - and criticism 
because we did not discuss his work exhaustively - demand that the astronomical 
issues be raised again.

The archaeological material was abundant, but could not provide precise dates.  
Astronomical data are one of the few means of securing potentially relevant and 
precise dates; astronomical sources are rare in the ancient Near East.  
Huber (1999/2000a : 55) described the very Ur III omens we selected as being 
"so detailed and unsystematic that they appear to contain actual records of 
observations."  This statement includes two separate judgments about the 
material.  One is that the data are detailed and the other that the organization 
of the material implied that actual observations were recorded.

The primary evaluation of the information in the omens depends upon their 
reliability.  This means that the more reliable sources must be selected, and that 
preference be given to the most important parameters.  For me, "reliable sources" 
include astronomical data that can be extracted and which are unambiguous in 
character; it does not mean that the sources are "reliable informants."  
An astronomer can recognize whether the data recorded reflect unique phenomena 
which can be isolated.  An astronomer cannot judge the level of confidence which 
can be assigned to any given text: only colleagues from the relevant field are 
qualified to translate the texts, thus allowing others to make and debate judgments 
of this kind.

Some omens seem to meet these sets of standards.  We exploited the above-mentioned 
methodology for the analysis of the records with astronomical content in the 
EAE series, essentially that discussed by Brinkman (2000), with the goal of securing
 information which may be limited in quantity, but of potentially high quality.  
Even if the conclusions are indecisive from the standpoint of the questions originally 
posed, they may still allow certain conclusions to be excluded by demonstrating 
discrepancies between the alleged records and the actual events.

EAE 20 and 21 describe eclipses which have been linked to the Ur III period.  
They have been regarded as being more reliable than the Venus Tablet.  By 
their very nature records of eclipses differ significantly from the records of the 
visibility of the planet Venus.  The omina that were interpreted as recording an 
eclipse are regarded as reliable precisely because the description should allow 
he eclipses to be identified, if the description preserved is correct.  By contrast, 
it is a well-known fact that the Venus Tablet dataset contains numerous errors.  
Separating the parameters from the background noise must be accomplished by 
a statistical evaluation to secure the non-corrupted information.  The problem is 
not new, since statistical methods inevitably face two fundamental 
obstacles:\footnote{For modern methods of the study of non-linear astrophysical problems, 
from planetary dynamics to data analysis, see e.g. Gurzadyan and Kocharyan 1994, 
and Gurzadyan and Ruffini 2000.}

(a)	the arbitrary character of the choice of the sample parameters to be 
statistically weighed;

(b)	the arbitrary character of the normalization of probabilities.

Each of these factors is crucial and a slight variation of the arbitrary parameters can 
suffice to distort the data to such an extent that the conclusions are completely 
unreliable.

There are, however, also purely astronomical reasons which render the alleged 
records of eclipses more valuable than the Venus Tablet.  Doubt can be cast upon 
the records of the Venus Tablet concerning issues ranging from their apparent 
regularity to local conditions influencing visibility.  This means that even if the 
data preserved were completely uncorrupted, the interpretation of the Venus 
Tablet would still be far from straightforward.

\section{UR  III  ECLIPSES}

The informative data recorded for the two lunar eclipses of EAE 20 and 21 can be 
reduced to 4 parameters :

(a)	the entering and 

(b)	exit positions of the darkening of the lunar disk;

(c)	the watch-times of the beginning and 

(d)	the end of the eclipse.

Some data recorded, namely the day of the eclipses (EAE 20 is dated to 14 Simanu 
and EAE 21 to 14 Addaru) and the direction of the wind are not informative for our 
purpose.

In Gasche et al. (1998) we proposed the eclipse of 27 June 1954 BC as a candidate 
for the EAE 20 description.  In his critical analysis of our conclusions, Koch (1998)
concluded that "the eclipse of June 27, 1954 BC would appear to fit the omen 
EAE 20 IIIB exactly."  He based this on the fact that the correlation fits the exit 
position ("lower west").  Since the eclipse began below or close to the horizon in 
the twilight, "the beginning of the darkening was nothing more than the assumed 
position of the lunar disk and therefore not binding for the examination of the 
Ur III Simanu eclipse."  In contrast to the precise relationship between our candidate 
and interpretation, Koch concluded that the eclipse proposed by Huber (1987a; 
25 July 2095 BC) contradicted the textual evidence referring to 
the "lower west" although the exit of the actual eclipse must have been the 
"upper west."

The crucial role of the end of the darkening results from the simple fact that the 
end of a sequence is always more reliably recorded than its beginning.  The 
observer must adjust his position or instruments (in later periods) and thus the 
description of the close of the sequence is always more accurate.  Regardless of all 
other errors - including scribal errors - the exit information has to be considered 
more reliable for an eclipse identification that the entry information.  Huber (1987a) 
overlooked this basic astronomical reality and assumed the contrary, with the 
unfortunate result that his proposed eclipse of 2095 BC fails to fulfil this significant 
criterion.  The record of the beginning of the event will also be even more defective 
if the eclipse began below the horizon, so that the beginning was in fact never 
observed, but merely reconstructed.

In these cases, the duration and time may also be relevant.  The crucial criterion of
EAE 20 IIIA that "the evening watch passes and the middle watch is touched" or 
of EAE 20 IIIB, that "the evening watch is half over," is more suitable for the eclipse 
of 1954 BC than the eclipse of 2095 BC (Koch 1998).

The eclipse of EAE 20 IIIB also meets an additional condition, namely that "the lunar 
eclipse must end at the time when the weakly shining stars become visible" (Donbaz 
and Koch 1995, 71; Koch 1998), which is again suitable for the eclipse of 1954 BC, 
but not that of 2095 BC.

The situation with the EAE 21 XII eclipse is more or less identical, but with one 
critical difference, the possible presence of a scribal error or misinterpretation.  
The crucial parameter of the textual evidence concerning the exit position of the
darkening ("north") matches the eclipse of 17 March 1912 BC precisely, but not 
that of 13 April 2053 BC (Huber 1987a) ("west").  The only ambiguity in the eclipse 
of 1912 BC is the duration of the watch-times.  The omen records an impossibly 
long eclipse far exceeding any possible eclipse, as Koch (1998) has pointed out.

In the final analysis, however, the eclipses of 1954 BC and 1912 BC (proposed in 
Gasche et al. 1998) fit the most significant information in EAE 20 and 21 while the 
eclipses of 2095 BC and 2053 BC (proposed by Huber et al. 1982, 
Huber 1987a, 1987b, 1999/2000b, and Huber, this volume) fail to meet the criteria, 
and actually contradict the conditions of the omina.  On this basis, at least, they
 should be excluded, regardless of any statistical evaluations.

This illustrates one of the dangers of conclusions based on statistical evaluations.  
Neglecting the basic astronomical information, Huber (1987b) found that eclipses -
which were ruled out for astronomical reasons - reflected a state where
"the probability of error is below 1 per cent".
Statistics cannot usefully be employed to 
determine the probability of something which is excluded in advance.  Similar 
difficulties lie behind other statements such :  "even if the chronology is wrong, 
we will find a matching eclipse in a 20-year window" (1999/2000a : 56), if the 
proper input information is wrong.

Huber (1999/2000a : 61) remarked that he had excluded the pair of eclipses proposed 
by us because he assumed that one of them began "too early."  As we saw above, 
this is not the case :  the eclipses correspond to the crucial factor of the exit and 
are otherwise largely compatible with the givens (Koch 1998), Huber had 
subjectively reduced the size of his sample before commencing his statistical project, 
and thereby endangered any potential reliability his methods and conclusions could 
have had.

\section{THE  VENUS  TABLET}

As mentioned by Huber and others, the data from the Venus Tablet has been 
thrown into doubt due to the numerous errors which have corrupted the text.  
It is therefore useful to examine the reliability of the presence of the 56/64 year 
Venus cycle preserved in the text, and whether this can be used.

The Old Babylonian calendar was not continually modified, but was altered only 
when the discrepancy between the seasons and the calendar forced the issue 
(Neugebauer 1975).  One cannot expect that if certain Venus visibility data fit one 
given 8-year-period, that they would therefore necessarily fit the period of 56 or 64 
years before/after with the same (a) probability and (b) accuracy.

The probability implies that if a primary period is found to fit the data with a certain 
probability, then the 56/64 year earlier or later periods cannot have the same 
probability.  The accuracy of any given record indicates the time separating the 
record from the last revision of the lunar calendar.  This means that, for example, 
6 or 9 Venus cycles, i.e. 48 or 72 years respectively, may happen to fit the primary 
period that would be required by the lunar calendar, since the scribe was obviously
 recording the official date recognized in the kingdom and not intent on making 
precise astronomical measurements.

Changes in the calendar appear inexplicable without specific information, and such
 changes are very remarkable when all other data appears to be absolutely constant.  
For example, a historian studying modern records based on a small data sample and
 only the knowledge of the Julian and Gregorian calendars, would be perplexed by 
the apparent mismatch of events in England in the 18th century or in Russia in the
 early 20th, unless a sudden 13-day shift might be recognized.

The meaning of such difficulties is overlapped with the intrinsic noise in the Venus 
Tablet, compounded by visibility aspects, e.g. when the first/last visibility records 
might actually be made later/earlier than it is required by the limiting stellar 
magnitude (Reiner and Pingree 1975), and potential scribal errors, quite aside 
from unconscious restorations in antiquity.  Altogether, this means that the 
visibility records could actually be assigned completely different positions 
in the cycle.  This demonstrates that the 20 chronologies proposed by 
Huber et al. (1982; Huber 1987a) cannot be used in astronomical historical 
arguments.

The reliability of the lunar calendar in the Venus Tablet is further undermined by 
the nature of certain errors.  For example, the text of Omen 11 mentions the last 
visibility date III 11 and an interval of 9 months 4 days.  For the same period, 
Omens 21 and 59 record XII 11 an interval of 4 days (Reiner and Pingree 1975), 
thus indicating the adjustment between the date and the interval made by the 
copyist.  This reflected the tradition of inserting "appropriate data" rather than 
actual data drawn from observations, in order to secure regularity (Neugebauer 
1983; see also Fatoohi et al. 1999, 51).

What can, therefore, be extracted from the Venus Tablet with a high degree of
 confidence ?  In Gasche et al. (1998), we advocated the use of the least noisy 
signature, namely, the use of the basic 8-year cycle to the exclusion of all others, 
i.e., any trace of the 584 day Venus synodic period (roughly 5 synodic Venus 
periods = 8 sidereal years minus 4 days).  Only the relative sequence of the 
inferior and superior conjunctions is reflected in the tablet and not the absolute 
lunar calendar.  In stating that the recorded observations were not visible for the 
proposed inferior conjunctions compatible for our proposed year for
 Ammisaduqa 1, Huber (1999/2000a : 53) shows that he has failed to 
appreciate the significance of our method.  Since our system does not recognize the 
validity of Huber's calendar for 20 chronologies, our 8-year Venus cycles are not 
anchored as Huber's are.  The result is that the 8-year Venus cycle data are 
compatible with the data drawn from the lunar eclipses, because they do not 
rely on Huber's 20 chronologies, which are based on the rejected 56/64 year 
cycles.  The background noise in the Venus Tablet makes this virtually the 
only logical approach : recognize the 8-year cycles.

Later, I became aware that Hunger (2000) has also recognized only the 8-year 
cycle.

The relative sequence of the inferior and superior conjunctions and not the lunar 
calendar lie at the base of our approach to the 8-year Venus cycles.  This sequence 
was unique and its signature would have been the last to have been effaced and lost
as a result of careless copying and inattentive observation.  The date for 
Ammisaduqa 1 = 1550 BC is thus anchored by the lunar eclipses but not by the 
lunar calendar.

To illustrate the significance and value of the method, the Venus data for 
Ammisaduqa 1 are calculated for 1702, 1646, 1582 and 1550 BC, corresponding to 
the traditional High, Middle, Low Chronologies, and Ultra-Low Chronology 
(Gasche et al. 1998).  In order to avoid the possible impact of local visibility 
conditions and other uncertainties, the elongation alone is presented.  The 
data thus relies on the angular separation of Venus from the Sun, by means 
of which Venus becomes invisible from a specific critical angle $\Theta_{cr}$.  The 
glare critical angle for average vision is given by the formula (Schaefer 1991) :
$$
\log \Theta_{cr} = 0.2 (9.28 + m_v),
$$
where $m_v$ is the visual stellar magnitude.  The Table 1 below includes the dates of 
Venus's passage according to the angle $\Theta_{cr} = 11^{\circ}$.  

\begin{table}                 
{
\renewcommand{\tabcolsep}{2.2mm}
{\footnotesize
\begin{center}
\begin{tabular}{ccccccccc}
\multicolumn{9}{c}{\bf T a b l e\qquad 1}\\[3mm]
\hline
\hline\\[-3.5mm]
Venus& Date& $m_{\rm v}$& $\Theta_{\rm cr}$& \qquad &
Venus& Date& $m_{\rm v}$& $\Theta_{\rm cr}$\\
\hline\\[-3mm]
        &              &     &                &  &
        &              &     &                \\
        & {\bf 1703 BC}&     &                &  &
        & {\bf 1647 BC}&     &                \\
Southern& 14.08        & -4.1& $11^o51^\prime$&  &
Southern& 28.07        & -4.1& $11^o11^\prime$\\
Southern& 15.08        & -4.1& $10^o55^\prime$&  &
Southern& 29.07        & -4.0& $10^o13^\prime$\\
Southern& 23.08        & -4.2& $10^o42^\prime$&  &
Southern& 06.08        & -4.1& $10^o14^\prime$\\
Southern& 24.08        & -4.2& $11^o36^\prime$&  &
Southern& 07.08        & -4.1& $11^o12^\prime$\\
        &              &     &                &  &
        &              &     &                \\
        & {\bf 1702 BC}&     &                &  &
        & {\bf 1646 BC}&     &                \\
Southern& 29.04        & -3.9& $11^o14^\prime$&  &
Southern& 14.04        & -3.9& $11^o02^\prime$\\
Southern& 30.04        & -3.9& $10^o58^\prime$&  &
Southern& 15.04        & -3.9& $10^o46^\prime$\\
Northern& 17.07        & -3.9& $10^o43^\prime$&  &
Northern& 02.07        & -3.9& $10^o53^\prime$\\
Northern& 18.07        & -3.9& $11^o00^\prime$&  &
Northern& 03.07        & -3.9& $11^o10^\prime$\\
        &              &     &                &  &
        &              &     &                \\
        & {\bf 1701 BC}&     &                &  &
        & {\bf 1645 BC}&     &                \\
Northern& 22.03        & -4.0& $12^o03^\prime$&  &
Northern& 07.03        & -4.0& $11^o16^\prime$\\
Northern& 23.03        & -4.0& $10^o50^\prime$&  &
Northern& 08.03        & -4.0& $10^o14^\prime$\\
Northern& 03.04        & -4.0& $10^o59^\prime$&  &
Northern& 17.03        & -4.1& $10^o54^\prime$\\
Northern& 04.04        & -4.1& $12^o12^\prime$&  &
Northern& 18.03        & -4.1& $12^o00^\prime$\\
Northern& 01.12        & -3.9& $11^o13^\prime$&  &
Northern& 13.11        & -3.9& $11^o11^\prime$\\
Northern& 02.12        & -3.9& $10^o58^\prime$&  &
Northern& 14.11        & -3.9& $10^o56^\prime$\\
\hline
        &              &     &                &  &
        &              &     &                \\
        &              &     &                &  &
        &              &     &                \\
        & {\bf 1583 BC}&     &                &  &
        & {\bf 1551 BC}&     &                \\
Southern& 08.07        & -4.0& $11^o03^\prime$&  &
Southern& 28.06        & -4.0& $11^o12^\prime$\\
Southern& 09.07        & -4.0& $09^o53^\prime$&  &
Southern& 29.06        & -4.0& $09^o56^\prime$\\
Southern& 19.07        & -4.1& $10^o24^\prime$&  &
Southern& 10.07        & -4.1& $10^o24^\prime$\\
Southern& 20.07        & -4.1& $11^o37^\prime$&  &
Southern& 11.07        & -4.1& $11^o42^\prime$\\
        &              &     &                &  &
        &              &     &                \\
        & {\bf 1582 BC}&     &                &  &
        & {\bf 1550 BC}&     &                \\
Southern& 26.03        & -3.9& $11^o14^\prime$&  &
Southern& 17.03        & -3.9& $11^o10^\prime$\\
Southern& 27.03        & -3.9& $10^o59^\prime$&  &
Southern& 18.03        & -3.9& $10^o54^\prime$\\
Northern& 14.06        & -3.9& $10^o47^\prime$&  &
Northern& 05.06        & -3.9& $10^o44^\prime$\\
Northern& 15.06        & -3.9& $11^o04^\prime$&  &
Northern& 06.06        & -3.9& $11^o00^\prime$\\
        &              &     &                &  &
        &              &     &                \\
        & {\bf 1581 BC}&     &                &  &
        & {\bf 1549 BC}&     &                \\
Northern& 17.02        & -4.1& $11^o55^\prime$&  &
Northern& 08.02        & -4.1& $11^o25^\prime$\\
Northern& 18.02        & -4.1& $10^o54^\prime$&  &
Northern& 09.02        & -4.1& $10^o28^\prime$\\
Northern& 26.02        & -4.1& $10^o16^\prime$&  &
Northern& 17.02        & -4.1& $10^o44^\prime$\\
Northern& 27.02        & -4.1& $11^o12^\prime$&  &
Northern& 18.02        & -4.1& $11^o43^\prime$\\
Northern& 24.10        & -3.9& $11^o04^\prime$&  &
Northern& 14.10        & -3.9& $11^o03^\prime$\\
Northern& 25.10        & -3.9& $10^o49^\prime$&  &
Northern& 15.10        & -3.9& $10^o48^\prime$\\
\hline
\end{tabular}
\end{center}}}
\end{table}

The relative magnitudes for each chronology are 
calculated and Universal Time is set at 00:41.  The calculations were performed
 with the Ephemeris - Tool 4.1 Version 4.1003, 2000 software, elaborated by 
M. Dings based on Newcomb theories, which is quite sufficient for this purpose.  
The sequence of the conjunctions can be clearly followed for each chronology.  
The data in the Venus Tablet can only be linked to any given chronology if 
anchored by an alternative link.  Without any reliable data (such as the Ur III 
eclipses), the signal cannot be anchored with confidence.

\section{MONTH  LENGTHS}

The principal of the method is to fit the distribution of 29 and 30 day months to 
any given chronology.  With this goal, Huber (1987b) relied upon assumptions 
which could possibly distort the conclusions, e.g.

(a)	assuming that data from incompatible collections, e.g. Late and 
Neo-Babylonian, share the same month-length distribution and data quality, 
which is certainly not fulfilled; 

(b)	the distribution of month lengths for his datasets cannot be given 
binomial or Poissonian distribution, due to potential inhomogeneities in the 
frequencies;

(c)	the use of sample-depended parameters.

Let us illustrate the situation with his analysis of Ur III month lengths where an 
explicit plot is available (Huber 1987b; Exhibit 4).  The data have $K_0$ = 83 
discrepancies from N = 228 month lengths.  He concludes that the result 
favours the High Chronology while conceding "that the statistical significance 
is not quite reached."  This does not quite compare with the confidence he has 
since expressed on the very same results in his later papers.
The situation is, however, more complicated, as it revolves around the issue of 
the choice of the correctness of one of the two distributions.  A proper evaluation 
of A (correct chronology) and B (best of wrong chronologies) must also include the 
null hypothesis.  Then, at $K > K_0$, the area between the distribution curve of A and 
the abscissa (discrepancy) axis will correspond to the significance level of the 
hypothesis "A is correct" but the criterion is excluding it.  A $K < K_0$, the area below 
the distribution curve corresponding to the hypothesis B indicates the probability 
of the error of the second category, i.e., when hypothesis B is correct but the 
criterion accepts the hypothesis A.  Since $K_0$ is sample dependent and 
non-homogeneous and is hence unstable, no conclusion can be drawn on the 
correctness of either of the hypotheses (e.g., Prokhorov and Statulevicuius 2000). 
 As Schaefer (1999) noted, "without a disproof of the null hypothesis, Occam's 
Razor would imply that the claimed discovery is false."
The situation is similar or even worse for his other samples : 6-13 discrepancies 
in 21 month lengths (Ammisaduqa) and 20 discrepancies in 57 
(Hammurabi-Samsuiluna) (Huber 1987b).

Besides this, there are also crucial astronomical-observational effects which were 
neglected by Huber.  First, the lunar synodic period is not 29.53, but can vary 
from 29.27 to 29.83 days (see Schaefer 1992).  Second, as is noted in Schaefer (1992), 
Huber had not taken into account the seasonal variations of the extinction 
coefficients, while there are numerous effects to be taken into account, i.e. the 
lunar albedo, the relative positions of the moon with respect to the sun and the
 horizon, the dependence of the extinction coefficient on the date, latitude, 
humidity, on the optical pathlength in the atmosphere, the atmospheric 
refraction, the brightness of the twilight sky, the detection probability of the 
human eye, etc.

Such statistical approaches cannot, therefore, be used for any far-reaching 
conclusions concerning any chronologies.  The month-lengths, however, may 
serve as a useful complementary tool for more complete and reliable datasets.

\section{CONCLUSION}

We have, therefore, observed that Huber (1987a; 1987b) arrived at erroneous
eclipse identifications while failing to consider others which were potentially 
valid.  This was at least partially due to his neglect of important bits of 
astronomical information recorded in the omens.  The same neglect also 
accounts for his claims for the potential eclipses proposed for the Akkadian 
period (Huber 1999/2000a).\footnote{There are many cases which 
demonstrate that the purely statistical-probability approach is insufficient and
 misleading.  Nobel laureate Piotr Kapitza drew attention to the fact that 
"numerous mistakes and false discoveries [are] obtained via statistical analysis, 
when the scientist aims to obtain the desirable result" (Kapitza 1981).}  
Huber's interpretation of the Venus Tablet is at 
least partially illusory since (a) the estimations of confidence levels are biased, 
(b) the 56/64 year cycles cannot be traced through and linked with a moving 
lunar calendar, and (c) his month-length assumptions are not statistically 
significant enough to match his chronological aims.

Without these supports, the High Chronology lacks any astronomical basis.  
Nor does any actual alternative support exist for the High Chronology (Hunger, 
personal communication).  Huber (1999/2000b) himself concedes that the High 
Chronology is not supported by other alternative data.  He nevertheless continues 
to claim that his astronomical arguments remain valid, even though - as we saw 
above - there are reasonable grounds for disputing them.

The analysis of astronomical records by distinguishing the reliability of their 
information as well as their possible chronological significance promises to 
produce results.  Isolating the eclipse parameters of EAE 20 and 21 and 
extracting the limited data from the Venus Tablet allow a high degree of 
confidence in employing these as the astronomical anchor for the Ultra-Low 
Chronology proposed in Gasche et al. (1998).\footnote{Due to technical reasons I had no possibility to see the final galley
 proofs of Gasche et al. (1998), as well as the artwork in Gurzadyan (2000) prepared
 by the editors, and therefore here I tried to maximally clarify the points which
 might be misinterpreted from there.}

As stated at the outset, the astronomical and historical sources were sought to 
buttress an argument which was primarily based on archaeological evidence.  
The typological and stratigraphical evidence originally presented has not been 
thrown into doubt.  Scholars not associated with the development of this 
hypothesis studying seals (e.g., Gualandi 1998), Egyptian materials (Krauss 
1998 and personal communication), and Anatolian dendrochronological evidence 
(Kuniholm et al. 1996) have all provided additional interpretations of evidence 
compatible with the Ultra-Low Chronology.  Other contributors present at the 
Ghent Colloquium (e.g. the contributions of Gates, van Soldt and Warburton in 
this volume) likewise presented interpretations compatible with the Ultra-Low 
Chronology.

While no compelling arguments have been brought against any elements of the 
astronomical or archaeological evidence, Hunger (2000) has likewise come to 
share the opinion of the Venus data, suggesting that only the 8-year cycles can be 
taken as reliable markers.

The recent conclusion drawn by Spence on the shortening of the 
Egyptian chronology of the IIIrd millennium BC by means of her 
remarkable idea on the alignment of the pyramids at Giza
 (Spence 2000, Gingerich 2000) is also compatible (Warburton, in this volume) 
with the Ultra-Low Chronology.

Our work was built on that of previous generations of scholars and has been 
achieved through intensive co-operation between the representatives of each 
discipline.  We believe that such collaboration can lead to a deeper insight 
into analogous multi-scale problems, and believe that it will become a common 
practice in the future.\footnote{The news that a record of another solar eclipse linked to the life of the
 Assyrian King Samsi-Adad has been discovered suggests such further co-operation.  
Particularly, the eclipses of 8 October -1763 and 28 July -1732 would have been 
visible at Mari, but only with a co-operation that the potential chronological 
significance of these and other eclipses can be established.}

\renewcommand{\baselinestretch}{1}
\section*{\it References}

\begin{description}
\item[ ]\hspace{-1mm}   J.A.Brinkman 2000, Panel talk at the Intern.
    Colloquium on Ancient Near Eastern Chronology (Ghent, July, 2000).
    \\[-7mm]
\item[ ]\hspace{-1mm}  V.Donbaz and J.Koch 1995, Ein Astrolab der
    dritten Generation, Nv.10, {\it JCS}, 47, 71. \\[-7mm]
\item[ ]\hspace{-1mm}  L.J.Fatoohi, F.R.Stephenson and S.S.Al-
    Dargazelli 1999, The Babylonian First Visibility of the Lunar
    Crescent: Data and Criterion, {\it Journ.Hist.Astr.} 30, 51.
    \\[-7mm]
\item[ ]\hspace{-1mm}  H.Gasche, J.A.Armstrong, S.W.Cole S.W. and
    V.G.Gurzadyan 1998,  {\it Dating the Fall of Babylon. A Reappraisal
    of Second-Millennium Chronology}, Mesopotamian History and
    Environment, Series II, Memoirs IV (Ghent and Chicago). \\[-7mm]
\item[]\hspace{-1mm}  O.Gingerich 2000, Plotting the Pyramids, {\it Nature}, 408, 297.
\item[ ]\hspace{-1mm}  G.Gualandi 1998,  Terqa Glyptic Data Highly
    Support a Low Chronology, {\it NABU}, 4, 133. \\[-7mm]
\item[ ]\hspace{-1mm}  V.G.Gurzadyan and A.A.Kocharyan 1994, {\it
    Paradigms of the Large-Scale Universe}, Gordon and Breach (New
    York). \\[-7mm]
\item[ ]\hspace{-1mm}  V.G.Gurzadyan 2000, Astronomy and the Fall
    of Babylon, {\it Sky and Telescope,} 100, 40. \\[-7mm]
\item[ ]\hspace{-1mm}  V.G.Gurzadyan and R.Ruffini (eds.) 2000,
    {\it The Chaotic Universe}, World Scientific (New York). \\[-7mm]
\item[ ]\hspace{-1mm}  P.J.Huber with collaboration of A.Sachs,
    M.Stol, R.W.Whiting, E.Leichty, C.B.F.Walker and C. van Driel 1982,
    Astronomical dating of Babylon I and Ur III (G.Buccelati, ed.)
    (Malibu). \\[-7mm]
\item[]\hspace{-1mm}  P.J.Huber, 1987a, Dating by Lunar Eclipse
    Omens with Speculations on the Birth of Omen Astrology, in: From
    Ancient Omens to Statistical Mechanics (J.L.Berggren and
    B.R.Goldstein, eds.), University Library (Copenhagen). \\[-7mm]
\item[]\hspace{-1mm}  P.J.Huber, 1987b, Astronomical Evidence
    for the Long and against the Middle and Short Chronologies, in:
    {\it High, Middle or Low?} (P.Astrom, ed.),  (Gothenburg). \\[-7mm]
\item[]\hspace{-1mm}  P.J.Huber 1999/2000a, Astronomical dating of Ur III
and Akkad, {\it Archiv fur Orientforschung}, 46-47, 50.  \\[-7mm]
\item[]\hspace{-1mm}  P.J.Huber 1999/2000b, Review of Gasche et al 1998.
{\it Archiv  fur Orientforschung}, 46-47, 287.\\[-7mm]
\item[]\hspace{-1mm}  H.Hunger 2000 Talk at the Intern.
    Colloquium on Ancient Near Eastern Chronology (Ghent, July, 2000). \\[-7mm]
\item[]\hspace{-1mm}  P.Kapitza 1981, {\it Experiment, Theory,
    Practice}, Nauka (Moscow). \\[-7mm]
\item[]\hspace{-1mm}  J.Koch 1998, Neues von den Ur III-
    Mondenklipsen, {\it NABU}, 4,  126. \\[-7mm]
\item[]\hspace{-1mm}  R.Krauss 1998, Altaegyptische Sirius-und
    Monddaten aus dem 19. und 18. Jahrhundert vor Christi Geburt
    (Berliner Illahun Archiv), in: Aegypten und Levante/Egypt and the
    Levant. {\it International Journal for Egyptian Archaeology and
    Related Disciplines}, 8, 113. \\[-7mm]
\item[]\hspace{-1mm}  P.I.Kuniholm et al 1996, Anatolian tree
    rings and the absolute chronology of the Eastern Mediterranean,
    2220-718 BC, {\it Nature}, 381, 780.  \\[-7mm]
\item[]\hspace{-1mm}  O.Neugebauer 1975, {\it A History of
    Ancient Mathematical Astronomy}, Springer-Verlag (New York). \\[-7mm]
\item[]\hspace{-1mm}  Yu.V.Prokhorov and V.Statulevicius 2000,
    {\it Limit Theorems of Probability Theory}, Springer-Verlag (New
    York). \\[-7mm]
\item[]\hspace{-1mm}  E.Reiner and D.Pingree 1975, {\it
    Babylonian Planetary Omens. Part I. The Venus Tablet of
    Ammisaduqa}, (Malibu). \\[-7mm]
\item[]\hspace{-1mm}  B.E.Schaefer  1991, Eclipse Earthshine, {\it
    Pub.Astr.Soc.Pacific.}, 103, 645. \\[-7mm]
\item[]\hspace{-1mm}  B.E.Schaefer  1992, The Length of the Lunar Month,
     {\it Archaeastronomy}, 17, S32. \\[-7mm]
\item[]\hspace{-1mm}  B.E.Schaefer 1999, New Methods and
    Techniques for Historical Astronomy and Archaeo-astronomy; talk at
    Oxford VI and SAEC conference on Astronomy and Cultural Diversity,
    (La Laguna, June, 1999). \\[-7mm]
\item[]\hspace{-1mm}  K.Spence 2000, Ancient Egyptian Chronology and the     Astronomical Orientation of Pyramids, {\it Nature}, 408, 320. \\

\end{description}

\end{document}